\documentclass[twocolumn,showpacs,preprintnumbers]{revtex4}
\usepackage{amsmath}
\usepackage{amssymb}
\usepackage{amsfonts}
\usepackage{graphicx}
\usepackage{dcolumn}
\usepackage{bm}

\input{tcilatex}

\begin{document}

\title{{EGT through Quantum Mechanics }\\
{\& from Statistical Physics to Economics}}
\author{Esteban Guevara Hidalgo$^{\dag \ddag }$}
\affiliation{$^{\dag }$Departamento de F\'{\i}sica, Escuela Polit\'{e}cnica Nacional,
Quito, Ecuador\\
$^{\ddag }$SI\'{O}N, Autopista General Rumi\~{n}ahui, Urbanizaci\'{o}n Ed%
\'{e}n del Valle, Sector 5, Calle 1 y Calle A \# 79, Quito, Ecuador}

\begin{abstract}
By analyzing the relationships between a socioeconomical system modeled
through evolutionary game theory and a physical system modeled through
quantum mechanics we show how although both systems are described through
two theories apparently different both are analogous and thus exactly
equivalents. The extensions of quantum mechanics to statistical physics and
information theory let us use some of their definitions for the best
understanding of the behavior of economics and biology. The quantum analogue
of the replicator dynamics is the von Neumann equation. A system in where
all its members are in Nash equilibrium is equivalent to a system in a
maximum entropy state. Nature is a game in where its players compete for a
common welfare and the equilibrium of the system that they are members. They
act as a whole besides individuals like they obey a rule in where they
prefer to work for the welfare of the collective besides the individual
welfare.
\end{abstract}

\pacs{03.65.-w, 02.50.Le, 03.67.-a, 89.65.Gh}
\maketitle
\email{esteban\_guevarah@yahoo.es}

\section{Introduction}

Why has it been possible to apply some methods of statistical physics to
economics and biology? It is a good reason to say that physics is a model
which tries to describe phenomena and behaviors and if this model fits and
describes almost perfectly the observed and the measured even in the
economic world then there is no problem or impediment to apply physics to
solve problems in economics and biology. But, could economics, biology and
statistical physics be correlated? Could it have a relationship between
quantum mechanics and game theory? or could quantum mechanics even enclose
theories like games and the evolutionary dynamics? This possibility could
make quantum mechanics a theory more general that we have thought.

\bigskip

Problems in economy and finance have attracted the interest of statistical
physicists. Ausloos et al \cite{1} analyzed fundamental problems pertain to
the existence or not long-, medium-, short-range power-law correlations in
economic systems as well as to the presence of financial cycles. They
recalled methods like the extended detrended fluctuation analysis and the
multi-affine analysis emphasizing their value in sorting out correlation
ranges and predictability. They also indicated the possibility of crash
predictions. They showed the well known financial analyst technique, the so
called moving average, to raise questions about fractional Brownian motion
properties. The $(m,k)$-Zipf method and the $i-$variability diagram
technique were presented for sorting out short range correlations.\bigskip

J.-P. Bouchaud \cite{2} analyzed three main themes in the field of
statistical finance, also called econophysics: (i) empirical studies and the
discovery of universal features in the statistical texture of financial time
series, (ii) the use of these empirical results to devise better models \ of
risk and derivative pricing, of direct interest for the financial industry,
and (iii) the study of \textquotedblleft agent-based
models\textquotedblright\ \ in order to unveil the basic mechanisms that are
responsible for the statistical\ \ \textquotedblleft
anomalies\textquotedblright\ \ observed in financial time series.\bigskip

Statistical physicists are also extremely interested in fluctuations \cite{3}%
. One reason physicists might want to quantify economic fluctuations is in
order to help our world financial system avoid \textquotedblleft economic
earthquakes\textquotedblright . Also it is suggested that in the field of
turbulence, we may find some crossover with certain aspects of financial
markets.

\bigskip

Kobelev et al \cite{4} used methods of statistical physics of open systems
for describing the time dependence of economic characteristics (income,
profit, cost, supply, currency, etc.) and their correlations with each
other. They also offered nonlinear equations (analogies of known
reaction-diffusion, kinetic, Langevin equation) to describe appearance of
bifurcations, self-sustained oscillational processes, self-organizations in
economic phenomena.

\bigskip

It is generally accepted that entropy can be used for the study of economic
systems consisting of large number of components \cite{5}. I. Antoniou et al 
\cite{6} introduced a new approach for the presentation of economic systems
with a small number of components as a statistical system described by
density functions and entropy. This analysis is based on a Lorenz diagram
and its interpolation by a continuous function. Conservation of entropy in
time may indicate the absence of macroscopic changes in redistribution of
resources. Assuming the absence of macro-changes in economic systems and in
related additional expenses of resources, we may consider the entropy as an
indicator of efficiency of the resources distribution. This approach is not
limited by the number of components of the economic system and can be
applied to wide class of economic problems. They think that the bridge
between distribution of resources and proposed probability distributions may
permit us to use the methods of nonequilibrium statistical mechanics for the
study and forecast of the dynamics of complex economic systems and to make
correct management decisions.

\bigskip

Statistical mechanics and economics study big ensembles: collections of
atoms or economic agents, respectively. The fundamental law of equilibrium
statistical mechanics is the Boltzmann-Gibbs law, which states that the
probability distribution of energy $E$ is $P(E)=Ce^{-E/T}$, where $T$ is the
temperature, and $C$ is a normalizing constant. The main ingredient that is
essential for the derivation of the Boltzmann-Gibbs law is the conservation
of energy. Thus, one may generalize that any conserved quantity in a big
statistical system should have an exponential probability distribution in
equilibrium \cite{7}. In a closed economic system, money is conserved. Thus,
by analogy with energy, the equilibrium probability distribution of money
must follow the exponential Boltzmann-Gibbs law characterized by an
effective temperature equal to the average amount of money per economic
agent. Dr\u{a}gulescu and Yakovenko demonstrated how the Boltzmann-Gibbs
distribution emerges in computer simulations of economic models. They
considered a thermal machine, in which the difference of temperature allows
one to extract a monetary profit. They also discussed the role of debt, and
models with broken time-reversal symmetry for which the Boltzmann-Gibbs law
does not hold.

\bigskip

Recently the insurance market, which is one of the important branches of
economy, have attracted the attention of physicists \cite{8}. Some concepts
of the statistical mechanics, specially the maximum entropy principle is
used for pricing the insurance. Darooneh obtained the price density based on
this principle, applied it to multi agents model of insurance market and
derived the utility function. The main assumption in his work is the
correspondence between the concept of the equilibrium in physics and
economics. He proved that economic equilibrium can be viewed as an
asymptotic approximation to physical equilibrium and some difficulties with
mechanical picture of the equilibrium may be improved by considering the
statistical description of it. Tops{\scriptsize \O }e \cite{9} also has
suggested that thermodynamical equilibrium equals game theoretical
equilibrium.

\bigskip

In this paper we try to find a deeper relationship between quantum mechanics
and game theory. By using some elements of evolution that are contained
inside evolutionary game theory through the replicator dynamics, game theory
is generalized. Evolutionary game theory is compared with quantum mechanics
and it is shown how although both systems analyzed are described through two
theories apparently different both are analogous and thus exactly
equivalents. So, we could make use of some of their concepts, laws and
definitions for the best understanding of the behavior of economics and
biology. The quantum analogue of the replicator dynamics is the von Neumann
equation. A system in where all its members are in Nash equilibrium is
equivalent to a system in a maximum entropy state. Nature is a game in where
its players act as a whole besides individuals like they obey a rule in
where they prefer to work for the welfare of the collective besides the
individual welfare. Should a socioeconomical system also satisfy a
Collective Welfare Principle?

\section{Classical, Evolutionary \& Quantum Games}

Game theory \cite{10,11,12} is the study of decision making of competing
agents in some conflict situation. It tries to understand the birth and the
development of conflicting or cooperative behaviors among a group of
individuals who behave rationally and strategically according to their
personal interests. Each member in the group strive to maximize its welfare
by choosing the best courses of strategies from a cooperative or individual
point of view.

\bigskip

The central equilibrium concept in game theory is the Nash Equilibrium. A 
\textit{Nash equilibrium }(NE) is a set of strategies, one for each player,
such that no player has an incentive to unilaterally change his action.
Players are in equilibrium if a change in strategies by any one of them
would lead that player to earn less than if he remained with his current
strategy. A \textit{Nash equilibrium} satisfies the following condition%
\begin{equation}
E(p,p)\geq E(r,p)\text{,}  \label{1}
\end{equation}%
where $E(s_{i},s_{j})$ is a real number that represents the payoff obtained
by a player who plays the strategy $s_{i}$\ against an opponent who plays
the strategy $s_{j}$. A player can not increase his payoff if he decides to
play the strategy $r$ instead of $p.$

\bigskip

Evolutionary game theory \cite{13,14,15} does not rely on rational
assumptions but on the idea that the Darwinian process of natural selection 
\cite{16} drives organisms towards the optimization of reproductive success 
\cite{17}. Instead of working out the optimal strategy, the different
phenotypes in a population are associated with the basic strategies that are
shaped by trial and error by a process of natural selection or learning. The
natural selection process that determines how populations playing specific
strategies evolve is known as the replicator dynamics \cite{18,14,15,19}
whose stable fixed points are Nash equilibria \cite{11}. The central
equilibrium concept of evolutionary game theory is the notion of
Evolutionary Stable Strategy introduced by J. Smith and G. Price \cite{20,13}%
. An ESS is described as a strategy which has the property that if all the
members of a population adopt it, no mutant strategy could invade the
population under the influence of natural selection. ESS are interpreted as
stable results of processes of natural selection.

\bigskip

Consider a large population in which a two person game $G=(S,E)$ is played
by randomly matched pairs of animals generation after generation. Let $p$ be
the strategy played by the vast majority of the population, and let $r$ be
the strategy of a mutant present in small frequency. Both $p$ and $r$ can be
pure or mixed. An \textit{evolutionary stable strategy} (ESS) $p$ of a
symmetric two-person game $G=(S,E)$ is a pure or mixed strategy for $G$
which satisfies the following two conditions%
\begin{gather}
E(p,p)>E(r,p)\text{,}  \notag \\
\text{If }E(p,p)=E(r,p)\text{ then }E(p,r)>E(r,r)\text{.}  \label{2}
\end{gather}%
Since the stability condition only concerns to alternative best replies, $p$
is always evolutionarily stable if $(p,p)$ is a strict equilibrium point. An
ESS is also a Nash equilibrium since is the best reply to itself and the
game is symmetric. The set of all the strategies that are ESS is a subset of
the NE of the game. A population which plays an ESS can withstand an
invasion by a small group of mutants playing a different strategy. It means
that if a few individuals which play a different strategy are introduced
into a population in an ESS, the evolutionarily selection process would
eventually eliminate the invaders.

\bigskip

Quantum games have proposed a new point of view for the solution of the
classical problems and dilemmas in game theory. Quantum games are more
efficient than classical games and provide a saturated upper bound for this
efficiency \cite{21,22,23,24,25,26}.

\section{Replicator Dynamics \& EGT}

The model used in EGT is the following: Each agent in a n-player game where
the $i^{th}$ player has as strategy space $S_{i}$ is modeled by a population
of players which have to be partitioned into groups. Individuals in the same
group would all play the same strategy. Randomly we make play the members of
the subpopulations against each other. The subpopulations that perform the
best will grow and those that do not will shrink and eventually will vanish.
The process of natural selection assures survival of the best players at the
expense of the others. The natural selection process that determines how
populations playing specific strategies evolve is known as the replicator
dynamics%
\begin{gather}
\frac{dx_{i}}{dt}=\left[ f_{i}(x)-\left\langle f(x)\right\rangle \right]
x_{i}\text{,}  \label{3} \\
\frac{dx_{i}}{dt}=\left[ \sum_{j=1}^{n}a_{ij}x_{j}-%
\sum_{k,l=1}^{n}a_{kl}x_{k}x_{l}\right] x_{i}\text{.}  \label{4}
\end{gather}%
The probability of playing certain strategy or the relative frequency of
individuals using that strategy is denoted by frequency $x_{i}$. The fitness
function $f_{i}=\sum_{j=1}^{n}a_{ij}x_{j}$ specifies how successful each
subpopulation is, $\left\langle f(x)\right\rangle
=\sum_{k,l=1}^{n}a_{kl}x_{k}x_{l}$ is the average fitness of the population,
and $a_{ij}$ are the elements of the payoff matrix $A$. The replicator
dynamics rewards strategies that outperform the average by increasing their
frequency, and penalizes poorly performing strategies by decreasing their
frequency. The stable fixed points of the replicator dynamics are Nash
equilibria, it means that if a population reaches a state which is a Nash
equilibrium, it will remain there.

\bigskip

We can represent the replicator dynamics in matrix form 
\begin{equation}
\frac{dX}{dt}=G+G^{T}\text{.}  \label{5}
\end{equation}%
The relative frequencies matrix $X$ has as elements%
\begin{equation}
x_{ij}=\left( x_{i}x_{j}\right) ^{1/2}  \label{6}
\end{equation}%
and%
\begin{eqnarray}
\left( G+G^{T}\right) _{ij} &=&\frac{1}{2}\sum_{k=1}^{n}a_{ik}x_{k}x_{ij} 
\notag \\
&&+\frac{1}{2}\sum_{k=1}^{n}a_{jk}x_{k}x_{ji}  \notag \\
&&-\sum_{k,l=1}^{n}a_{kl}x_{k}x_{l}x_{ij}  \label{7}
\end{eqnarray}%
are the elements of the matrix $\left( G+G^{T}\right) $.

\bigskip

From this matrix representation we can find a Lax representation of the
replicator dynamics \cite{27}%
\begin{gather}
\frac{dX}{dt}=\left[ \left[ Q,X\right] ,X\right] \text{,}  \label{8} \\
\frac{dX}{dt}=\left[ \Lambda ,X\right] \text{.}  \label{9}
\end{gather}%
The matrix $\Lambda $ is equal to $\Lambda =\left[ Q,X\right] $ with%
\begin{equation}
(\Lambda )_{ij}=\frac{1}{2}\left[ \left( \sum_{k=1}^{n}a_{ik}x_{k}\right)
x_{ij}-x_{ji}\left( \sum_{k=1}^{n}a_{jk}x_{k}\right) \right]  \label{10}
\end{equation}%
and $Q$ is a diagonal matrix which has as elements%
\begin{equation}
q_{ii}=\frac{1}{2}\sum_{k=1}^{n}a_{ik}x_{k}.  \label{11}
\end{equation}

\bigskip

\section{Relationships between Quantum Mechanics \& Game Theory}

In table 1 we compare some characteristic aspects of quantum mechanics and
game theory.

{\scriptsize Table 1}

\begin{center}
\begin{tabular}{cc}
\hline
{\scriptsize Quantum Mechanics} & {\scriptsize Game Theory} \\ \hline
{\scriptsize n system members} & {\scriptsize n players} \\ 
{\scriptsize Quantum states} & {\scriptsize Strategies} \\ 
{\scriptsize Density operator} & {\scriptsize Relative frequencies vector}
\\ 
{\scriptsize Von Neumann equation} & {\scriptsize Replicator Dynamics} \\ 
{\scriptsize Von Neumann entropy} & {\scriptsize Shannon entropy} \\ 
{\scriptsize System Equilibrium} & {\scriptsize Payoff} \\ 
{\scriptsize Maximum entropy} & {\scriptsize Maximum payoff} \\ 
{\scriptsize \textquotedblleft Altruism\textquotedblright } & {\scriptsize %
Altruism or selfish } \\ 
{\scriptsize Collective Welfare principle} & {\scriptsize Minority Welfare
principle} \\ 
&  \\ \hline
\end{tabular}
\end{center}

It is easy to realize the clear resemblances and apparent differences
between both theories and between the properties both enjoy. This was a
motivation to try to find an actual relationship between both systems.

\bigskip

We have to remember that Schr\"{o}dinger equation describes only the
evolution of pure states in quantum mechanics. To describe correctly a
statistical mixture of states it is necessary the introduction of the
density operator%
\begin{equation}
\rho (t)=\sum_{i=1}^{n}p_{i}\left\vert \Psi _{i}(t)\right\rangle
\left\langle \Psi _{i}(t)\right\vert  \label{12}
\end{equation}%
which contains all the information of the statistical system. The time
evolution of the density operator is given by the von Neumann equation%
\begin{equation}
i\hbar \frac{d\rho }{dt}=\left[ \hat{H},\rho \right]  \label{13}
\end{equation}%
which is only a generalization of the Schr\"{o}dinger equation and the
quantum analogue of Liouville's theorem.

\bigskip

Evolutionary game theory has been applied to the solution of games from a
different perspective. Through the replicator dynamics it is possible to
solve not only evolutionary but also classical games. That is why EGT has
been considered like a generalization of classical game theory. The
bonestone of EGT is the concept of evolutionary stable strategy (ESS) that
is a strengthened notion of Nash equilibrium. The evolution of relative
frequencies in a population is given by the replicator dynamics. In a recent
work we showed that vectorial equation can be represented in a matrix
commutative form (\ref{9}). This matrix commutative form follows the same
dynamic than the von Neumann equation and the properties of its
correspondent elements (matrixes) are similar, being the properties
corresponding to our quantum system more general than the classical system.

\bigskip

The next table shows some specific resemblances between quantum statistical
mechanics and evolutionary game theory.

{\scriptsize Table 2}

\begin{center}
\begin{tabular}{cc}
\hline
{\scriptsize Quantum Statistical Mechanics} & {\scriptsize Evolutionary Game
Theory} \\ \hline
{\scriptsize n system members} & {\scriptsize n population members} \\ 
{\scriptsize Each member in the state }$\left\vert \Psi _{k}\right\rangle $
& {\scriptsize Each member plays strategy }$s_{i}$ \\ 
$\left\vert \Psi _{k}\right\rangle $ {\scriptsize with} $p_{k}\rightarrow $
\ $\rho _{ij}${\scriptsize \ } & $s_{i}${\scriptsize \ }$\ \ \rightarrow $%
{\scriptsize \ }$\ \ x_{i}$ \\ 
$\rho ,$ $\ \ \tsum_{i}\rho _{ii}{\scriptsize =1}$ & ${\scriptsize X,}$%
{\scriptsize \ \ }$\tsum_{i}x_{i}{\scriptsize =1}$ \\ 
${\scriptsize i\hbar }\frac{d\rho }{dt}{\scriptsize =}\left[ \hat{H},\rho %
\right] $ & $\frac{dX}{dt}{\scriptsize =}\left[ \Lambda ,X\right] $ \\ 
${\scriptsize S=-Tr}\left\{ {\scriptsize \rho }\ln {\scriptsize \rho }%
\right\} $ & ${\scriptsize H=-}\tsum\nolimits_{i}{\scriptsize x}_{i}\ln 
{\scriptsize x}_{i}$ \\ 
&  \\ \hline
\end{tabular}
\end{center}

Both systems are composed by $n$ members (particles, subsystems, players,
states, etc.). Each member is described by a state or a strategy which has
assigned a determined probability. The quantum mechanical system is
described by the density operator $\rho $ whose elements represent the
system average probability of being in a determined state. For evolutionary
game theory, we defined a relative frequencies matrix $X$ to describe the
system whose elements can represent the frequency of players playing a
determined strategy. The evolution of the density operator is described by
the von Neumann equation which is a generalization of the Schr\"{o}dinger
equation. While the evolution of the relative frequencies in a population is
described by the Lax form of the replicator dynamics which is a
generalization of the replicator dynamics in vectorial form.

\bigskip

In table 3 we show the properties of the matrixes $\rho $ and $X$.

{\scriptsize Table 3}

\begin{center}
$%
\begin{tabular}{cc}
\hline
{\scriptsize Density Operator} & {\scriptsize Relative freq. Matrix} \\ 
\hline
$\rho ${\scriptsize \ is Hermitian} & $X${\scriptsize \ is Hermitian} \\ 
${\scriptsize Tr\rho (t)=1}$ & ${\scriptsize TrX=1}$ \\ 
${\scriptsize \rho }^{2}{\scriptsize (t)\leqslant \rho (t)}$ & ${\scriptsize %
X}^{2}{\scriptsize =X}$ \\ 
${\scriptsize Tr\rho }^{2}{\scriptsize (t)\leqslant 1}$ & ${\scriptsize TrX}%
^{2}{\scriptsize (t)=1}$ \\ 
&  \\ \hline
\end{tabular}%
$
\end{center}

Although both systems analyzed are described through two apparently
different theories (quantum mechanics and game theory), both systems are
analogous and thus exactly equivalents \cite{28,29}. So, we can take some
concepts, laws and definitions from quantum mechanics and physics for the
best understanding of the behavior of economics and biology. Quantum
mechanics could be used to explain more correctly biological and economical
processes and even it could encloses theories like games and evolutionary
dynamics \cite{27,28,29}. This could make quantum mechanics a more general
theory that we had thought. The resemblances between both systems and the
similarity in the properties of their corresponding elements let us to
define and propose the next quantization relationships.

\section{Quantum Replicator Dynamics \& Quantization Relationships}

Let us propose the next quantization relationships%
\begin{gather}
x_{i}\rightarrow \sum_{k=1}^{n}\left\langle i\left\vert \Psi _{k}\right.
\right\rangle p_{k}\left\langle \Psi _{k}\left\vert i\right. \right\rangle
=\rho _{ii}\text{,}  \notag \\
(x_{i}x_{j})^{1/2}\rightarrow \sum_{k=1}^{n}\left\langle i\left\vert \Psi
_{k}\right. \right\rangle p_{k}\left\langle \Psi _{k}\left\vert j\right.
\right\rangle =\rho _{ij}\text{.}  \label{14}
\end{gather}%
A population will be represented by a quantum system in which each
subpopulation playing strategy $s_{i}$ will be represented by a pure
ensemble in the state $\left\vert \Psi _{k}(t)\right\rangle $ and with
probability $p_{k}$. The probability $x_{i}$ of playing strategy $s_{i}$ or
the relative frequency of the individuals using strategy $s_{i}$ in that
population will be represented as the probability $\rho _{ii}$ of finding
each pure ensemble in the state $\left\vert i\right\rangle $ \cite{27}.

\bigskip

Through these quantization relationships the replicator dynamics (in matrix
commutative form) takes the form of the equation of evolution of mixed
states. And also%
\begin{gather}
X\longrightarrow \rho \text{,}  \label{15} \\
\Lambda \longrightarrow -\frac{i}{\hbar }\hat{H}\text{,}  \label{16}
\end{gather}%
where $\hat{H}$ is the Hamiltonian of the physical system.

\bigskip

The equation of evolution of mixed states from quantum statistical mechanics
(\ref{13}) is the quantum analogue of \ the replicator dynamics in matrix
commutative form (\ref{9}) and both systems and their respective matrixes
have similar properties. Through these relationships we could describe
classical, evolutionary and quantum games and also the biological systems
that were described before through evolutionary dynamics with the replicator
dynamics.

\section{Games from QIT}

Lets consider a system composed by $N$ members, players, strategies, states,
etc. This system is described completely through certain density operator $%
\rho $ (\ref{12}), its evolution equation (the von Neumann equation) (\ref%
{13}) and its entropy. Classically, the system is described through the
matrix of relative frequencies $X$, the replicator dynamics and the Shannon
entropy. For the quantum case we define the von Neumann entropy as \cite%
{30,31,32,33}%
\begin{equation}
S=-Tr\left\{ \rho \ln \rho \right\} \text{,}  \label{17}
\end{equation}%
and for the classical case 
\begin{equation}
H=-\sum_{i=1}x_{ii}\ln x_{ii}  \label{18}
\end{equation}%
which is the Shannon entropy over the relative frequencies vector $x$ ( the
diagonal elements of $X$).

\bigskip

The time evolution equation of $H$ assuming that $x$ evolves following the
replicator dynamics is \cite{28,29,31} 
\begin{equation}
\frac{dH(t)}{dt}=Tr\left\{ U(\tilde{H}-X)\right\} \text{.}  \label{19}
\end{equation}%
$\tilde{H}$ is a diagonal matrix whose trace is equal to the Shannon entropy
and its elements are $\tilde{h}_{ii}=-x_{i}\ln x_{i}$. The matrix $U$ has as
elements $U_{i}=\sum_{j=1}^{n}a_{ij}x_{j}-\sum_{k,l=1}^{n}a_{kl}x_{l}x_{k}$.

\bigskip

With the purpose of calculating the time evolution of the von\ Neumann
entropy we approximate the logarithm of $\rho $ by series \cite{28,29,31} $%
\ln \rho =(\rho -I)-\frac{1}{2}(\rho -I)^{2}+\frac{1}{3}(\rho -I)^{3}$...
and 
\begin{eqnarray}
\frac{dS(t)}{dt} &=&\frac{11}{6}\tsum\limits_{i}\frac{d\rho _{ii}}{dt} 
\notag \\
&&-6\tsum\limits_{i,j}\rho _{ij}\frac{d\rho _{ji}}{dt}  \notag \\
&&+\frac{9}{2}\tsum\limits_{i,j,k}\rho _{ij}\rho _{jk}\frac{d\rho _{ki}}{dt}
\notag \\
&&-\frac{4}{3}\tsum\limits_{i,j,k,l}\rho _{ij}\rho _{jk}\rho _{kl}\frac{%
d\rho _{li}}{dt}+\zeta \text{.}  \label{20}
\end{eqnarray}

\bigskip

Entropy is the central concept of information theories. The von Neumann
entropy \cite{32,33} is the quantum analogue of Shannon's entropy but it
appeared 21 years before and generalizes Boltzmann's expression. Entropy%
\textbf{\ }in quantum information theory plays prominent roles in many
contexts, e.g., in studies of the classical capacity of a quantum channel 
\cite{34,35} and the compressibility of a quantum source \cite{36,37}.
Quantum information theory appears to be the basis for a proper
understanding of the emerging fields of quantum computation \cite{38,39},
quantum communication \cite{40,41}, and quantum cryptography \cite{42,43}.

\bigskip

In classical physics, information processing and communication is best
described by Shannon information theory. The Shannon entropy expresses the
average information we expect to gain on performing a probabilistic
experiment of a random variable which takes the values $s_{i}$ with the
respective probabilities $x_{i}$. It also can be seen as a measure of
uncertainty before we learn the value of that random variable. The Shannon
entropy of the probability distribution associated with the source gives the
minimal number of bits that are needed in order to store the information
produced by a source, in the sense that the produced string can later be
recovered.

\bigskip

We can define an entropy over a random variable $S^{A}$ (player's $A$
strategic space) which can take the values $\left\{ s_{i}^{A}\right\} $ (or $%
\left\{ \left\vert \Psi _{i}\right\rangle \right\} _{A}$) with the
respective probabilities $\left( x_{i}\right) _{A}$ (or $\left( \rho
_{ij}\right) _{A}$) \cite{31}. We could interpret the entropy of our game as
a measure of uncertainty before we learn what strategy player $A$ is going
to use. If we do not know what strategy a player is going to use every
strategy becomes equally probable and our uncertainty becomes maximum and
greater while greater is the number of strategies. If we would know the
relative frequency with player $A$ uses any strategy we can prepare our
reply in function of that most probable player $A$ strategy. Obviously our
uncertainty vanish if we are sure about the strategy our opponent is going
to use.

\bigskip

If player $B$ decides to play strategy $s_{j}^{B}$ against player $A$ which
plays the strategy $s_{i}^{A}$ our total uncertainty about the pair $(A,B)$
can be measured by an external \textquotedblleft referee\textquotedblright\
through the joint entropy of the system. This is smaller or at least equal
than the sum of the uncertainty about $A$ and the uncertainty about $B$. The
interaction and the correlation between $A$ and $B$ reduces the uncertainty
due to the sharing of information. The uncertainty decreases while more
systems interact jointly creating a new only system. If the same systems
interact in separated groups the uncertainty about them is bigger. We can
measure how much information $A$ and $B$ share and have an idea of how their
strategies or states are correlated by its mutual or correlation entropy. If
we know that $B$ decides to play strategy $s_{j}^{B}$ we can determinate the
uncertainty about $A$ through the conditional entropy.

\bigskip

Two external observers of the same game can measure the difference in their
perceptions about certain strategy space of the same player $A$ by its
relative entropy. Each of them could define a relative frequency vector and
the relative entropy over these two probability distributions is a measure
of its closeness. We could also suppose that $A$ could be in two possible
states i.e. we know that $A$ can play of two specific but different ways and
each way has its probability distribution (\textquotedblleft
state\textquotedblright ) that also is known. Suppose that this situation is
repeated exactly $N$ times or by $N$ people. We can made certain
\textquotedblleft measure\textquotedblright , experiment or
\textquotedblleft trick\textquotedblright\ to determine which the state of
the player is. The probability that these two states can be confused is
given by the classical or the quantum Sanov's theorem.

\section{Maximum Entropy \& the Collective Welfare Principle}

We can maximize $S$ by requiring that%
\begin{equation}
\delta S=-\sum_{i}\delta \rho _{ii}(\ln \rho _{ii}+1)=0  \label{21}
\end{equation}%
subject to the constrains $\delta Tr\left( \rho \right) =0$ and $\delta
\left\langle E\right\rangle =0$. By using Lagrange multipliers%
\begin{equation}
\sum_{i}\delta \rho _{ii}(\ln \rho _{ii}+\beta E_{i}+\gamma +1)=0  \label{22}
\end{equation}%
and the normalization condition $Tr(\rho )=1$ we find that 
\begin{equation}
\rho _{ii}=\frac{e^{-\beta E_{i}}}{\sum_{k}e^{-\beta E_{k}}}  \label{23}
\end{equation}%
which is the condition that the density operator and its elements must
satisfy to our system tends to maximize its entropy $S$. If we maximize $S$
without the internal energy constrain $\delta \left\langle E\right\rangle =0$
we obtain%
\begin{equation}
\rho _{ii}=\frac{1}{N}  \label{24}
\end{equation}%
which is the $\beta \rightarrow 0$\ limit (\textquotedblleft high -
temperature limit\textquotedblright ) in equation (\ref{23}) in where a
canonical ensemble becomes a completely random ensemble in which all energy
eigenstates are equally populated. In the opposite low - temperature limit $%
\beta \rightarrow \infty $ tell us that a canonical ensemble becomes a pure
ensemble where only the ground state is populated. The parameter $\beta $ is
related to the \textquotedblleft temperature\textquotedblright\ $\tau $ as
follows%
\begin{equation}
\beta =\frac{1}{\tau }\text{.}  \label{25}
\end{equation}

\bigskip

By replacing $\rho _{ii}$ obtained in the equation (\ref{23}) in the von
Neumann entropy we can rewrite it in function of the partition function $%
Z=\sum_{k}e^{-\beta E_{k}}$, $\beta $ and $\left\langle E\right\rangle $
through the next equation%
\begin{equation}
S=\ln Z+\beta \left\langle E\right\rangle \text{.}  \label{26}
\end{equation}%
It is easy to show that the next relationships for the energy of our system
are satisfied%
\begin{gather}
\left\langle E\right\rangle =-\frac{1}{Z}\frac{\partial Z}{\partial \beta }=-%
\frac{\partial \ln Z}{\partial \beta }\text{,}  \label{27} \\
\left\langle \Delta E^{2}\right\rangle =-\frac{\partial \left\langle
E\right\rangle }{\partial \beta }=-\frac{1}{\beta }\frac{\partial S}{%
\partial \beta }\text{.}  \label{28}
\end{gather}%
We can also analyze the variation of entropy with respect to the average
energy of the system%
\begin{gather}
\frac{\partial S}{\partial \left\langle E\right\rangle }=\frac{1}{\tau }%
\text{,}  \label{29} \\
\frac{\partial ^{2}S}{\partial \left\langle E\right\rangle ^{2}}=-\frac{1}{%
\tau ^{2}}\frac{\partial \tau }{\partial \left\langle E\right\rangle }
\label{30}
\end{gather}%
and with respect to the parameter $\beta $%
\begin{gather}
\frac{\partial S}{\partial \beta }=-\beta \left\langle \Delta
E^{2}\right\rangle \text{,}  \label{31} \\
\frac{\partial ^{2}S}{\partial \beta ^{2}}=\frac{\partial \left\langle
E\right\rangle }{\partial \beta }+\beta \frac{\partial ^{2}\left\langle
E\right\rangle }{\partial \beta ^{2}}\text{.}  \label{32}
\end{gather}

\bigskip

If our systems are analogous and thus exactly equivalents, our physical
equilibrium should be also absolutely equivalent to our socieconomical
equilibrium. If in an isolated system each of its accessible states do not
have the same probability, the system is not in equilibrium. The system will
vary and will evolution in time until it reaches the equilibrium state in
where the probability of finding the system in each of the accessible states
is the same. The system will find its more probable configuration in which
the number of accessible states is maximum and equally probable. The whole
system will vary and rearrange its state and the states of its ensembles
with the purpose of maximize its entropy and reach its maximum entropy
state. We could say that the purpose and maximum payoff of a quantum system
is its maximum entropy state. The system and its members will vary and
rearrange themselves to reach the best possible state for each of them which
is also the best possible state for the whole system. This can be seen like
a microscopical cooperation between quantum objects to improve its state
with the purpose of reaching or maintaining the equilibrium of the system.
All the members of our quantum system will play a game in which its maximum
payoff is the equilibrium of the system. The members of the system act as a
whole besides individuals like they obey a rule in where they prefer the
welfare of the collective over the welfare of the individual. This
equilibrium is represented in the maximum system entropy in where the system
\textquotedblleft resources\textquotedblright\ are fairly distributed over
its members. A system where its members are in Nash Equilibrium (or ESS) is
exactly equivalent to a system in a maximum entropy state \cite{8,9}. 
\textit{\textquotedblleft A system is stable only if it maximizes the
welfare of the collective above the welfare of the individual. If it is
maximized the welfare of the individual above the welfare of the collective
the system gets unstable and eventually it collapses\textquotedblright\ }%
(Collective Welfare Principle\ \cite{27,28,29,31}).

\bigskip

There exist tacit rules inside a system. These rules do not need to be
specified or clarified and search the system equilibrium under the
collective welfare principle. The other \textquotedblleft
prohibitive\textquotedblright\ and \textquotedblleft
repressive\textquotedblright\ rules are imposed over the system when one or
many of its members violate the collective welfare principle and search to
maximize its individual welfare at the expense of the group. Then it is
necessary to establish regulations over the system to try to reestablish the
broken natural order.

\bigskip

Fundamentally, we could distinguish three states in every system: minimum
entropy state, maximum entropy state, and when the system is tending to
whatever of these two states. The natural trend of a physical system is to
the maximum entropy state. The minimum entropy state is a characteristic of
a \textquotedblleft manipulated\textquotedblright\ system i.e. externally
controlled or imposed. The \textquotedblleft
globalization\textquotedblright\ process has a behavior exactly equivalent
to a physical system that is tending to a maximum entropy state. So, the
analysis shown in this paper would predict the apparition of big common
markets and strong common currencies as the characteristics of a system that
is tending to that maximum entropy state. The system will find its
equilibrium by decreasing its number of currencies and markets until it gets
a state characterized by only one common currency and only one common market.

\section{The Why of the Applicability of Statistical Physics to Economics}

Quantum mechanics could be a much more general theory that we had thought.
It could encloses theories like EGT and evolutionary dynamics and we could
explain through this theory biological and economical processes. From this
point of view many of the equations, concepts and its properties defined
quantically must be more general that its classical versions but they must
remain inside the foundations of the new quantum version. So, our quantum
equilibrium concept also must be more general than the one defined
classically.

\bigskip

In our model we represent a population by a quantum system in which each
subpopulation playing strategy $s_{i}$ is represented by a pure ensemble in
the state $\left\vert \Psi _{k}(t)\right\rangle $ and with probability $%
p_{k} $. The probability $x_{i}$ of playing strategy $s_{i}$ or the relative
frequency of the individuals using strategy $s_{i}$ in that population is
represented by the probability $\rho _{ii}$ of finding each pure ensemble in
the state $\left\vert i\right\rangle $. Through these quantization
relationships the replicator dynamics (in matrix commutative form) takes the
form of the equation of evolution of mixed states. The quantum analogue of
the relative frequencies matrix is the density operator. The relationships
between these two systems described by these two matrixes and their
evolution equations would let us analyze the entropy of our system through
the well known von Neumann entropy in the quantum case and by the Shannon
entropy in the classical case. The properties that these entropies enjoy
would let us analyze a \textquotedblleft game\textquotedblright\ from a
different point of view through information and a maximum or minimum entropy
criterion.

\bigskip

Every game can be described by a density operator, the von Neumann entropy
and the quantum replicator dynamics. The density operator is maybe the most
important tool in quantum mechanics. From the density operator we can
construct and obtain all the statistical information about our system. Also
we can develop the system in function of its information and analyze it
through information theories under a criterion of maximum or minimum
entropy. There exists a strong relationship between game theories,
statistical mechanics and information theory. The bonds between these
theories are the density operator and entropy.

\bigskip

It is important to remember that we are dealing with very general and
unspecific terms, definitions, and concepts like state, game and system. Due
to this, the theories that have been developed around these terms like
quantum mechanics, statistical physics, information theory and game theory
enjoy of this generality quality and could be applicable to model any system
depending on what we want to mean for game, state, or system. Objectively
these words can be and represent anything. Once we have defined what system
is in our model, we could try to understand what kind of \textquotedblleft
game\textquotedblright\ is developing between its members and how they
accommodate their \textquotedblleft states\textquotedblright\ in order to
get their objectives. This would let us visualize what temperature, energy
and entropy would represent in our specific system through the
relationships, properties and laws that were defined before when we
described a physical system.

\bigskip

Entropy can be defined over any random variable and can be maximized subject
to different constrains. In each case the result is the condition the system
must follow to maximize its entropy. Generally, this condition is a
probability distribution function. For the case analyzed in this paper, this
distribution function depends on certain parameter \textquotedblleft $\beta $%
\textquotedblright\ which is related inversely with the system
\textquotedblleft temperature\textquotedblright . Depending on what the
variable over which we want determinate its grade of order or disorder is we
can resolve if the best for the system is a state of maximum or minimum
entropy. If we would measure the order or disorder of our system over a
resources distribution variable the best state for that system is those in
where its resources are fairly distributed over its members which would
represent a state of maximum entropy. By the other hand, if we define an
entropy over an acceptation relative frequency of a presidential candidate
in a democratic process the best would represent a minimum entropy state
i.e. the acceptation of a candidate by the vast majority of the population.

\section{Conclusions}

By analyzing the relationships between the systems described through quantum
mechanics and game theory we showed that both systems are analogous and thus
exactly equivalents. The extensions of quantum mechanics to statistical
physics and information theory let us use some of their concepts and
definitions for the best understanding of the behavior of economics and
biology. Quantum mechanics could be a much more general theory that we had
thought. It could encloses theories like EGT and evolutionary dynamics.

\bigskip

Due to the generality of the terms state, game and system, quantum
mechanics, statistical physics, information theory and game theory enjoy
also of this generality quality and could be applicable to model any system
depending on what we want to mean for game, state, or system. Once we have
defined what the term system is in our model, we could try to understand
what kind of \textquotedblleft game\textquotedblright\ is developing between
its members and how they accommodate its \textquotedblleft
states\textquotedblright\ in order to get their objectives.

\bigskip

Entropy can be defined over any random variable and can be maximized subject
to different constrains. Depending on what the variable over which we want
determinate its grade of order or disorder is we can resolve if the best for
the system is its state of maximum or minimum entropy. A system can be
internally or externally controlled with the purpose of guide it to a state
of maximum or minimum entropy depending of the ambitions of the members that
compose it or the \textquotedblleft people\textquotedblright\ who control
it. Every \textquotedblleft game\textquotedblright\ could be described by a
density operator with its entropy equal to von Neumann's and its evolution
equation given by the quantum replicator dynamics which is the von Neumann
equation.

\bigskip

The quantum analogue of the replicator dynamics is the von Neumann equation.
A system in where all its members are in Nash equilibrium is equivalent to a
system in a maximum entropy state. Nature is a game in where its players
compete for a common welfare and the equilibrium of the system that they are
members. They act as a whole besides individuals like they obey a rule in
where they prefer to work for the welfare of the collective besides the
individual welfare. This equilibrium is represented in the maximum system
entropy in where the system \textquotedblleft resources\textquotedblright\
are fairly distributed over its members. A system is stable only if it
maximizes the welfare of the collective above the welfare of the individual.
If it is maximized the welfare of the individual above the welfare of the
collective the system gets unstable and eventually it collapses.


\begin{thebibliography}{99}
\bibitem{1} M. Ausloos, N. Vandewalle, Ph. Boveroux, A.Minguet, K. Ivanova,
Physica A \textbf{247}, 229-240 (1999).

\bibitem{2} J.-P. Bouchaud, \textit{An introduction to statistical finance},
Physica A \textbf{313}, 238-251 (2002).

\bibitem{3} H. Eugene Stanley, \textit{Exotic statistical physics:
Applications to biology, medicine, and economics}, Physica A \textbf{285,}
1-17 (2000).

\bibitem{4} L.Ya. Kobelev, O.L. Kobeleva, Ya.L. Kobelev, \textit{Is it
Possible to Describe Economical Phenomena by Methods of Statistical Physics
of Open Systems?}, physics/0005010.

\bibitem{5} Masanao Aoki, \textit{New approaches to Macroeconomics Modeling}
(Cambridge University Press, Cambridge, 1996).

\bibitem{6} I. Antoniou, V.V. Ivanov, Yu.L. Korolev, A.V. Kryanev, V.V.
Matokhin, Z. Suchanecki, Physica A \textbf{304}, 525-534 (2002).

\bibitem{7} A. Dr\u{a}gulescu and V. M. Yakovenko, \textit{Statistical
mechanics of money}, Eur. Phys. J. B \textbf{17}, 723-729 (2000).

\bibitem{8} A. Darooneh, Entropy \textbf{8[1],} 18-24 (2006).

\bibitem{9} F. Tops{\scriptsize \O }e, Information Theoretical Optimization
Techniques, Kybernetika \textbf{15},\textbf{\ }8-27 (1979). F. Tops%
{\scriptsize \O }e, Game theoretical equilibrium, maximum entropy and
minimum information discrimination, in \textit{Maximum Entropy and Bayesian
Methods} (A. Mohammad-Djafari and G. Demoments (eds.), pp. 15-23, Kluwer,
Dordrecht, 1993).

\bibitem{10} J. von Neumann and O. Morgenstern, \textit{The Theory of Games
and Economic Behavior} ( Princeton \ University Press, Princeton, 1947).

\bibitem{11} R. B. Myerson, \textit{Game Theory: An Analysis of Conflict}
(MIT Press, Cambridge, 1991).

\bibitem{12} M. A. Nowak and K. Sigmund, Nature \textbf{398}, 367 (1999).

\bibitem{13} J. M. Smith, \textit{Evolution and The Theory of Games}
(Cambridge University Press, Cambridge, UK, 1982).

\bibitem{14} J. Hofbauer and K. Sigmund, \textit{Evolutionary Games and
Replicator Dynamics} (Cambridge University Press, Cambridge, UK, 1998).

\bibitem{15} J. Weibul, \textit{Evolutionary Game Theory} (MIT Press,
Cambridge, MA, 1995).

\bibitem{16} R. A. Fisher, \textit{The Genetic Theory of Natural Selection}
(Oxford, Clarendon Press, 1930).

\bibitem{17} P. Hammerstein and R. Selten, \textit{Game Theory and
Evolutionary Biology} (Handbook of Game Theory. Vol 2. Elsevier B.V., 1994).

\bibitem{18} P. D. Taylor and L. B. Jonker , \textit{Evolutionary stable
strategies and game dynamics}, Mathematical Biosciences \textbf{40}, 145-156
(1978).

\bibitem{19} R. Cressman, \textit{The Stability Concept of Evolutionary Game
Theory: A Dynamic Approach} (Springer-Verlag, New York, 1992).

\bibitem{20} J. M. Smith and G. R. Price, \textit{The logic of animal
conflict}, Nature \textbf{246}, 15 (1973).

\bibitem{21} D. A. Meyer, Phys. Rev. Lett. \textbf{82}, 1052-1055 (1999).

\bibitem{22} J. Eisert, M. Wilkens and M. Lewenstein, Phys. Rev. Lett. 
\textbf{83}, 3077 (1999).

\bibitem{23} L. Marinatto and T. Weber, Phys. Lett. A \textbf{272}, 291
(2000).

\bibitem{24} A. P. Flitney and D. Abbott, Proc. R. Soc. (London) A \textbf{%
459}, 2463-74 (2003).

\bibitem{25} E. W. Piotrowski and J. Sladkowski, Int. J. Theor. Phys. 
\textbf{42}, 1089 (2003).

\bibitem{26} Azhar Iqbal, PhD thesis, Quaid-i-Azam University, 2004,
quant-ph/0503176.

\bibitem{27} E. Guevara H., \textit{Quantum Replicator Dynamics}, Physica A 
\textbf{369/2}, 393-407 (2006).

\bibitem{28} E. Guevara H., \textit{The Why of the applicability of
Statistical Physics to Economics}, physics/0609088.

\bibitem{29} E. Guevara H., \textit{Quantum Econophysics}, in Proceedings of
Quantum Interaction 2007, AAAI Spring Symposia Series, Stanford University,
Palo Alto, published by the American Association of Artificial Intelligence,
AAAI PRESS Technical Report\textbf{\ SS-07-08}, 158-165 (2007).

\bibitem{30} E. Guevara H., \textit{Introduction to the study of entropy in
Quantum Games}, quant-ph/0604170.

\bibitem{31} E. Guevara H., \textit{Quantum Games Entropy}, Physica A (to be
published), quant-ph/0606045.

\bibitem{32} J. von Neumann, \textit{Thermodynamik quantummechanischer
Gesamheiten}, G\"{o}tt. Nach. \textbf{1} 273-291(1927).

\bibitem{33} J. von Neumann, \textit{Mathematische Grundlagen der
Quantenmechanik} (Springer, Berlin, 1932)

\bibitem{34} B. Schumacher and M. D. Westmoreland, Phys. Rev. A \textbf{56},
131 (1997).

\bibitem{35} A. S. Holevo, IEEE Trans. Inf. Theory \textbf{44}, 269 (1998).

\bibitem{36} B. Schumacher, Phys. Rev. A \textbf{51}, 2738 (1995).

\bibitem{37} R. Jozsa and B. Schumacher, J. Mod. Opt. \textbf{41}, 2343
(1994).

\bibitem{38} C. H. Bennett and D. P. DiVincenzo, Nature \textbf{377}, 389
(1995).

\bibitem{39} D. P. DiVincenzo, Science \textbf{270}, 255 (1995).

\bibitem{40} C. H. Bennett and S. J. Wiesner, Phys. Rev. Lett. \textbf{69},
2881 (1992).

\bibitem{41} C. H. Bennett et al., Phys. Rev. Lett. \textbf{70}, 1895 (1993).

\bibitem{42} A. Ekert, Nature \textbf{358}, 14 (1992).

\bibitem{43} C. H. Bennett, G. Brassard, and N. D. Mermin, Phys. Rev. Lett. 
\textbf{68}, 557 (1992).
\end{thebibliography}
\end{document}